\documentclass[epjc3]{svjour3}

\usepackage[T1]{fontenc}
\usepackage[utf8]{inputenc}
\usepackage{booktabs}
\usepackage{amsmath,amssymb}
\usepackage{graphicx}
\usepackage{mathptmx}
\usepackage{flushend}
\usepackage[numbers,sort&compress]{natbib}
\usepackage[colorlinks,citecolor=blue,urlcolor=blue,linkcolor=blue]{hyperref}

\newcommand{\dif}{\mathrm{d}}

\newcommand{\imag}{\ensuremath{\mathrm{i}}}

\journalname{Eur. Phys. J. C}

\begin{document}

\title{Greybody factor and quasinormal modes of Regular Black Holes}

\subtitle{}

\author{Ángel Rincón\thanksref{e1,addr1} \and Victor Santos\thanksref{e2,addr2}}

\thankstext{e1}{e-mail: victor\_santos@fisica.ufc.br}
\thankstext{e2}{e-mail: angel.rincon@pucv.cl}

\institute{Instituto de Física, Pontificia Universidad Católica de Valparaíso, Avenida Brasil 2950, Casilla 4059, Valparaíso, Chile \label{addr1} \and Fundação Cearense de Apoio ao Desenvolvimento Científico e Tecnológico (FUNCAP), Av. Oliveira Paiva, 941, Cidade dos Funcionários, 60822-130, Fortaleza, Ceará, Brazil \label{addr2}}

\date{Received: date / Accepted: date}

\maketitle

\begin{abstract}
In this work, we investigate the quasinormal frequencies of a class of regular black hole solutions which generalize Bardeen and Hayward spacetimes. In particular, we analyze scalar, vector and gravitational perturbations of the black hole both with the semianalytic WKB method and their time-domain profiles. We analyze in detail the behavior of the spectrum depending on the parameter \(p/q\) of the black hole, the quantum number of angular momentum and the \(s\) number. In addition, we compare our results with the classical solution valid for \(p = q = 1\).
\end{abstract}

\tableofcontents

\section{Introduction}
\label{sec:introduction}

One of the most striking predictions of General Relativity (GR)~\cite{Einstein:1916vd} is the existence of Black Holes (BHs), objects which produce a region where not even light can escape. They can be formed in the extreme final stages of the gravitational collapse of stars. Such astrophysical objects are remarkably simple, being characterized by three parameters: mass, charge and angular momentum, by virtue of the so-called ``no-hair'' theorems~\cite{israel_event_1967,israel_event_1968,carter_axisymmetric_1971}. Since gravitational radiation emitted by an oscillating black hole can carry information about its inner properties like mass and charge~\cite{Hawking:1974rv,Hawking:1974sw}, this enables to use BHs as a laboratory for studying gravity in strong regimes, where quantum phenomena might take uttermost importance. Also, after Hawking’s seminal papers where the radiation from the black hole horizon was explained \cite{Hawking:1974sw,Hawking:1974rv}, black boles become an excellent laboratory to study and also enhance our comprehension about quantum gravity. Naively, the Hawking radiation is taken as black body radiation parametrized by the hawking temperature \(T_H\). However, the latter is just an approximated picture because emitted particles feel an effective potential barrier in the exterior region. Such barrier backscatters a percentage of the outgoing radiation back into the black hole~\cite{Kanti:2002nr}. Thus, the spectrum of the Hawking radiation as seen by an asymptotic observer has not a complete blackbody distribution: it is better described by a greybody distribution. The greybody factor can obtained from the transmission amplitude as the field modes pass from near horizon region to an asymptotic observer through the effective potential induced by the spacetime geometry. Estimation of this greybody factor is usually a difficult task and often one has to resort to approximations, usually in low/high frequency limits. There are monodromy methods~\cite{PhysRevD.93.024045,Castro_2013} and computations in a variety of scenarios~\cite{Gursel:2019fyd,Hyun:2019qgq,Kanzi:2019gtu,Chowdhury:2020bdi}, where one can also employ numerical approaches to estimate them~\cite{Gray:2015xig}. In particular, it should be noticed that a considerable part of the literature is dedicated to the case of black holes with singularities. In order to fix that problem, it is expected that quantum ingredients play a dominant role. Thus, although a complete theory of quantum gravity is still under construction, regular black hole solutions can be obtained utilizing additional matter, for instance, taking advantage of non-linear electrodynamics, as was previously pointed out by Ayon-Beato and Garcia in~\cite{AyonBeato:1998ub}. Additionally, an interesting feature of the gravitational collapse is that during the formation phase there is the emission of gravitational radiation, strong enough to be detected by the gravitational wave detectors~\cite{Kokkotas_Schmidt_1999}. This connection was strongly provided by the historical direct detection of GWs by LIGO, that two binary merger objects can coalesce into a super-massive black hole~\cite{Abbott:2016blz}. Moreover, recently was reported the first image of the super-massive black hole at the center of the giant elliptical galaxy Messier 87 (M87) by the Event Horizon Telescope~\cite{Akiyama:2019cqa,Akiyama:2019eap} which increased the interest on the physics behind this class of intriguing objects.

In the framework of GR, gravitational radiation arises as perturbations of spacetime itself. Due to the nonlinear nature of Einsteins equations, it is often very hard to find closed solutions, and usually one has to resort to perturbation theory. such perturbations give rise to a set of damped vibrations called \emph{quasinormal modes}~\cite{Konoplya_Zhidenko_2011}, complex numbers whose real part represents the actual frequency of the oscillation and the imaginary part represents the damping. A simple example of such oscillations are the oscillations of stars which are damped by internal friction~\cite{Kokkotas_Schutz_1992}. It is known that in general relativity damping might occur even in frictionless scenarios. This effect arises because energy may be radiated away towards infinity by gravitational waves. As it was previously pointed in Ref.~\cite{1971ApJ..170L.105P}, even linearized perturbations of black holes exhibit quasinormal modes. QNMs are also important in the investigation of the post-merger remnant of a binary black hole (BBH), as it coalescence settles to a Kerr black hole at a sufficiently late time after the merging phase. The stationary state is reached when the perturbed BH remnant emits gravitational waves (GWs) during a process known as the \emph{ringdown} (RD), and The GW corresponding to this phase is described by the linear perturbation theory of a Kerr BH~\cite{Vishveshwara_1970,Giesler_Isi_Scheel_Teukolsky_2019}.

As was previously said, the study of the quasinormal frequencies is quite relevant, because they encode information on how a black hole relaxes after it has been perturbed. Even more, such frequencies depend on i) the geometry and ii) the type of perturbations \cite{Flachi:2012nv}. Along the years, the study of QNM becomes more essential than ever, and certain seminal works have been performed up to now, for instance see \cite{Kokkotas:1999bd,Konoplya:2003ii,Konoplya:2011qq} and more recent works \cite{Barrau:2019swg,Konoplya:2019hlu,Panotopoulos:2017hns,Rincon:2018sgd,Destounis:2018utr,Rincon:2018ktz,Panotopoulos:2019qjk,Rincon:2020iwy,Rincon:2020pne,Panotopoulos:2020zbj,Oliveira:2018oha,Santos:2015gja,Devi:2020uac,Jusufi:2020agr,doi:10.1142/S021827181641008X}.

One issue of the classical description of gravity is the existence of spacetime singularities~\cite{hawking1975large}. Although the known solutions of Schwarzschild, Kerr and Reissner-Nordstrom have singularities protected by an event horizon~\cite{Penrose_2002}, the prediction that BHs emit radiation cause them to shrink until the singularity if reached. This is still a conundrum, as singularities are generally regarded as an indicative of the breakdown in the theory, requiring new Physics for a proper description. It is a common belief that only a consistent quantum theory of gravity could solve it properly~\cite{Frolov:1979tu}. Since there are no consistent models yet, phenomenological models have been proposed, and most of them are based on the avoidance of the central singularity. Such non-singular solutions are called \emph{regular black holes}~\cite{Hayward_2006,Flachi:2012nv,Toshmatov:2015wga}.

Given the increasing interest in gravitational wave astronomy and on QNMs of black holes, it would be interesting to investigate the QNM spectra from regular BHs. In previous works quasinormal modes of regular black holes were computed by several authors, see e.g.~\cite{Fernando_Correa_2012,Flachi_Lemos_2013,Toshmatov_Abdujabbarov_Stuchlik_Ahmedov_2015,Toshmatov_Stuchlik_Schee_Ahmedov_2018}. In this paper we propose to investigate the QNMs of a class of regular black holes which generalize Bardeen and Hayward spacetimes. In particular, we analyze scalar, vector and gravitational perturbations of the black hole both with the semianalytic WKB method and compute the greybody factor. Furthermore, it is essential to note that there is a vast collection of works where QNMs are calculated using the WKB approximation (see \cite{Panotopoulos:2017hns,Panotopoulos:2018pvu,Panotopoulos:2019qjk,Dey:2018cws,Wu:2017tfo,Yekta:2019por} and references therein).

The work is structured as follows. In section~\ref{sec:gravitational-bg-and-field-perturbations} we introduce the spacetime background of regular black holes and how to describe the field perturbations. In section~\ref{sec:qnms-in-wkb-approx} we compute the quasinormal frequencies emplying the 6th order WKB approximation, and in section~\ref{sec:gbf-computation} we compute the greybody factor. Section~\ref{sec:conclusion} is devoted to the conclusions.

\section{Gravitational background and field perturbations}
\label{sec:gravitational-bg-and-field-perturbations}

\subsection{Regular Black Holes}
\label{subsec:regular-bhs}

\begin{figure}[htbp]
\centering
\includegraphics[width=0.95\linewidth]{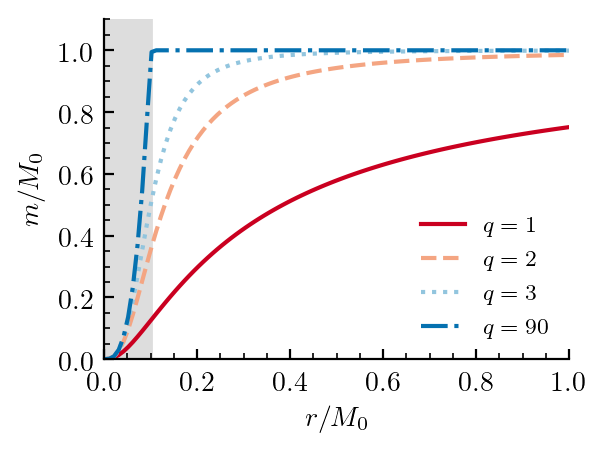}
\caption{\label{fig:mass-function}Mass function~\eqref{eq:mass-function} with \(p = 3\) and \(q > 0\). The parameter \(r_0\) is typically assumed to be microscopic (\(r_0 \ll M_0\)) and hence the exterior region of the BH can be very close to the Schwarzschild space-time. The \(q \to \infty\) case, as evidenced by the case \(q = 90\), corresponds to the usual matching between de Sitter and Schwarzchild solutions in the interior region of the BH. The grey shaded region indicates the interior region \(r < 2M_0\). In this graphic we have used \(r_0/M_0 = 10^{-1}\).}
\end{figure}

The class of regular BH solutions of interest in this work have the line element of a static, spherically symmetric black hole
\begin{equation}
\label{eq:line-element}
{\dif s}^2 = -f(r){\dif t}^2 + f(r)^{-1}{\dif r}^2 + r^2{\dif \Omega}^2,
\end{equation}
where \({\dif\Omega}^2\) is the metric of the 2-sphere and the lapse function \(f(r) = 1 -
2\,m(r)/r\) explicitly depends on the matter distribution. The particular mass function~\cite{Neves_Saa_2014}
\begin{equation}
\label{eq:mass-function}
m(r) = \frac{M_0}{{[1 + {(r_0/r)}^q]}^{p/q}},\ p,q\in\mathbb{Z}
\end{equation}
ensures an asymptotically flat spacetime for positive \(p\) and \(q\). \(M_0\) and \(r_0\) can be interpreted as mass and length parameters. The regular BH solutions proposed by Bardeen~\cite{ansoldi2008spherical} and Hayward~\cite{Hayward_2006} correspond to the choices \((p,q) = (3,2)\) and \((p,q) = (3,3)\) in equation~\eqref{eq:mass-function}. The limit of large \(r\) is
\begin{equation}
\label{eq:mass-function-large-r-limit}  
m(r) \approx M_0\bigg(1 - \frac{p}{q}{\bigg(\frac{r_0}{r}\bigg)}^q\bigg) ,
\end{equation}
the only restriction on~\eqref{eq:mass-function-large-r-limit} to obtain an asymptotically Schwarzschild solution is to have \(q > 0\). To compare our results with results found for Bardeen and Hayward solutions, we will restrict ourselves to the case \(p = 3\) and \(q > 0\). The behaviour of some typical cases is shown in Fig.~\eqref{fig:mass-function}. Thus, should be noticed that when the parameter \(q\) increases the mass function \(m(r)\) tends to its constant value \(M_0\).

\subsection{Field perturbations}
\label{subsec:field-perturbations}

\begin{figure}[htb]
\centering
\includegraphics[width=0.95\linewidth]{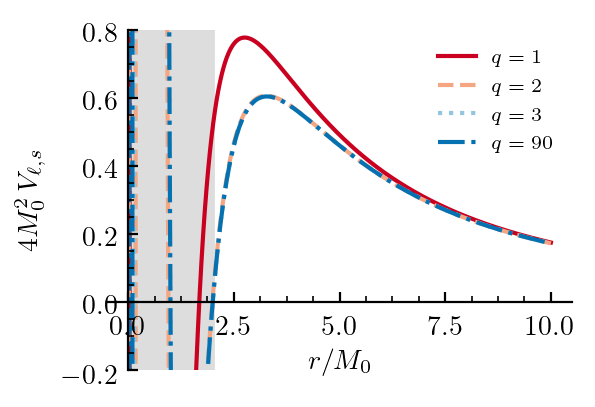}
\caption{\label{fig:rw-potential}Effective potential~\eqref{eq:rw-potential} for the mass function~\eqref{eq:mass-function} for gravitational perturbations \(s = 2\), multipole number \(\ell = 2\) and \(p = 3\) and \(q > 0\). The grey shaded region indicates the interior region \(r < 2M_0\). Notice that in the exterior region the potential is always positive and peaked.}
\end{figure}

\begin{figure}[htb]
\centering
\includegraphics[width=0.95\linewidth]{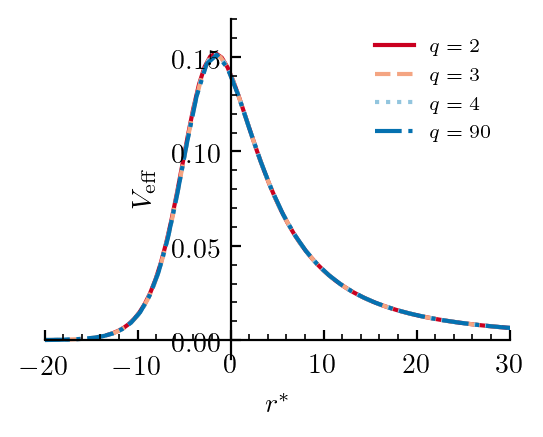}
\caption{\label{fig:rw-potential-tortoise}Effective potential~\eqref{eq:rw-potential} for the mass function~\eqref{eq:mass-function} for gravitational perturbations \(s = 2\), multipole number \(\ell = 2\) and \(p = 3\) and \(q > 1\) with respect to the tortoise coordinate \(r_\ast\).}
\end{figure}

This subsection is devoted to introduce the basic formalism as well as the theoretical grounds regarding perturbations in black holes in four dimensional spacetime. Thus, we will focus on scalar, electromagnetic and gravitational field perturbations in the fixed background given by Eq.~\eqref{eq:line-element} and mass function~\eqref{eq:mass-function}. Due to spherical symmetry, the fields can be decomposed into spherical harmonics, and the equations of motion can be reduced to the form
\begin{equation} \label{eq:rw-equation}
- \frac{\partial^2 \Psi}{\partial t^2}
+ \frac{\partial^2 \Psi}{\partial r_{\ast}^2} + V(r_{\ast})\Psi = 0,
\end{equation}
where the tortoise coordinate \(r_\ast\) is defined according to the differential equation
\begin{equation}
\label{eq:tortoise-coordinate}
\frac{\dif r_{\ast}}{\dif r} = \frac{1}{f(r)}.
\end{equation}
and the effective potential is
\begin{equation}
\label{eq:rw-potential}
V(r) = f(r)\bigg(\frac{\ell(\ell + 1)}{r^2} + \frac{1-s^2}{r}f^{\,\prime}(r)\bigg),
\end{equation}
where \(\ell\) denotes the multipole number of the spherical harmonics decomposition and \(s = 0, 1, 2\) is the spin of the perturbation. Its behavior is shown in Fig.~\ref{fig:rw-potential}.
From the behaviour showed in~\ref{fig:rw-potential} we can see that in the exterior region \(r > 2M_0\) the effective potential does not varies sensibly for \(q > 1\). Therefore, one should not expect large differences in the perturbation oscillation frequencies in the limit of large \(q\). This can be further inferred by observing the behavior of the effective potential with respect to the tortoise coordinate~\eqref{eq:tortoise-coordinate}, as shown in Fig.~\ref{fig:rw-potential-tortoise}.
Employing the stationary ansatz \(\Psi(t, r_{\ast}) \sim e^{\imag\omega t}\psi(r_{\ast})\), equation~\eqref{eq:rw-potential} becomes
\begin{equation}
\label{eq:schrodinger-qnm-equation}
\frac{\dif^2 \psi}{\dif r_{\ast}^2} + \big[\omega^2 - V(r_{\ast})\big]\psi = 0,
\end{equation}
which precisely takes the form of a Schrödinger-like equation. The frequencies of the temporal decomposition take the form \(\omega = \omega_R + \imag\omega_I\), where \(\omega_R\) is the oscillation frequency of the QNM and \(\omega_I\) is the damping time. Therefore, any mode with \(\omega_I < 0\) is unstable.

\section{QNMs of regular BHs in the WKB approximation}
\label{sec:qnms-in-wkb-approx}

The QNMs can be computed by imposing proper boundary conditions on~\eqref{eq:schrodinger-qnm-equation}, where the fields are purely ingoing at the BH horizon and purely outgoing at the spatial infinity. With such boundary conditions, the resulting frequencies are complex. If the potential~\eqref{eq:rw-potential} is peaked and falls to a constant in the asymptotic region, one can compute the QNM frequencies from~\eqref{eq:schrodinger-qnm-equation} employing the WKB approximation described by Schutz, Iyer and Will~\cite{PhysRevD.35.3632,1985ApJ291L33S,PhysRevD.35.3621} and posteriorly improved by Konoplya \cite{PhysRevD.68.024018}. With this last improvement the QNMs can be computed by
\begin{equation}
\frac{\imag Q_0}{\sqrt{2 Q''_0}} - \sum_{i = 2}^p \Lambda_i = n + \frac{1}{2},\quad n = 0, 1, \dots
\end{equation}
where the correction term \(\Lambda_i\) can be obtained for different orders of approximation. \(n\) is the overtone number and \(Q^{(i)}_0\) is the \(i\)-th derivative of \(Q = \omega^2 - V\) computed at the maximum of the potential. It is worth to mention that the accuracy of the WKB method is dependent of the multipole number and overtone: in general the approximation is appropriate for \(\ell > n\) and it is not applicable for \(\ell < n\). 

The results for the quasinormal modes are presented in tables \ref{tab:qnms-scalar-field}, \ref{tab:qnms-electromagnetic-field} and \ref{tab:qnms-gravitational-field}.

\begin{table}
\begin{tabular}{llllll}
\toprule
\(\ell\) & \(n\) & \(q = 2\) & \(q = 3\) & \(q = 4\) & \(q = 10\) \\
\midrule
 0 &  0 &  \(0.110750-0.100666\imag\) &  \(0.110475-0.100819\imag\) &  \(0.110467-0.100818\imag\) &  \(0.110467-0.100816\imag\) \\
 1 &  0 &  \(0.293430-0.097648\imag\) &  \(0.292921-0.097755\imag\) &  \(0.292910-0.097761\imag\) &  \(0.292910-0.097762\imag\) \\
 1 &  1 &  \(0.265209-0.306075\imag\) &  \(0.264491-0.306493\imag\) &  \(0.264472-0.306517\imag\) &  \(0.264471-0.306518\imag\) \\
 2 &  0 &  \(0.484470-0.096654\imag\) &  \(0.483660-0.096759\imag\) &  \(0.483642-0.096766\imag\) &  \(0.483642-0.096766\imag\) \\
 2 &  1 &  \(0.464821-0.295251\imag\) &  \(0.463871-0.295603\imag\) &  \(0.463847-0.295626\imag\) &  \(0.463847-0.295627\imag\) \\
 2 &  2 &  \(0.431610-0.507945\imag\) &  \(0.430418-0.508653\imag\) &  \(0.430387-0.508698\imag\) &  \(0.430386-0.508700\imag\) \\
\bottomrule
\end{tabular}
\caption{\label{tab:qnms-scalar-field}Quasinormal frequencies of scalar perturbations for \(p = 3\) and \(q > 1\). for the regular Black Hole where \(r_0/M_0 = 0.1\) and .}
\end{table}

\begin{table}
\begin{tabular}{llllll}
\toprule
\(\ell\) & \(n\) & \(q = 2\) & \(q = 3\) & \(q = 4\) & \(q = 10\) \\
\midrule
 1 &  0 &  \(0.248752-0.092567\imag\) &  \(0.248207-0.092631\imag\) &  \(0.248192-0.092637\imag\) &  \(0.248191-0.092637\imag\) \\
 1 &  1 &  \(0.215143-0.293803\imag\) &  \(0.214324-0.294097\imag\) &  \(0.214296-0.294117\imag\) &  \(0.214295-0.294118\imag\) \\
 2 &  0 &  \(0.458440-0.094907\imag\) &  \(0.457613-0.095004\imag\) &  \(0.457594-0.095011\imag\) &  \(0.457593-0.095011\imag\) \\
 2 &  1 &  \(0.437541-0.290373\imag\) &  \(0.436560-0.290704\imag\) &  \(0.436535-0.290727\imag\) &  \(0.436534-0.290728\imag\) \\
 2 &  2 &  \(0.402191-0.500999\imag\) &  \(0.400942-0.501681\imag\) &  \(0.400907-0.501726\imag\) &  \(0.400906-0.501728\imag\) \\
 3 &  0 &  \(0.658054-0.095510\imag\) &  \(0.656925-0.095610\imag\) &  \(0.656899-0.095617\imag\) &  \(0.656898-0.095617\imag\) \\
 3 &  1 &  \(0.643005-0.289388\imag\) &  \(0.641767-0.289708\imag\) &  \(0.641737-0.289730\imag\) &  \(0.641737-0.289731\imag\) \\
 3 &  2 &  \(0.615266-0.491421\imag\) &  \(0.613826-0.492022\imag\) &  \(0.613789-0.492062\imag\) &  \(0.613788-0.492064\imag\) \\
 3 &  3 &  \(0.579237-0.705454\imag\) &  \(0.577529-0.706432\imag\) &  \(0.577482-0.706495\imag\) &  \(0.577481-0.706498\imag\) \\
\bottomrule
\end{tabular}
\caption{\label{tab:qnms-electromagnetic-field}Quasinormal frequencies of electromagnetic perturbations for \(p = 3\) and \(q > 1\). for the regular Black Hole where \(r_0/M_0 = 0.1\) and .}
\end{table}

\begin{table}
\begin{tabular}{llllll}
\toprule
\(\ell\) & \(n\) & \(q = 2\) & \(q = 3\) & \(q = 4\) & \(q = 10\) \\
\midrule
 2 &  0 &  \(0.374511-0.088768\imag\) &  \(0.373643-0.088881\imag\) &  \(0.373620-0.088890\imag\) &  \(0.373619-0.088891\imag\) \\
 2 &  1 &  \(0.347558-0.273042\imag\) &  \(0.346340-0.273445\imag\) &  \(0.346298-0.273478\imag\) &  \(0.346297-0.273480\imag\) \\
 2 &  2 &  \(0.300515-0.476533\imag\) &  \(0.298601-0.477480\imag\) &  \(0.298523-0.477555\imag\) &  \(0.298520-0.477560\imag\) \\
 3 &  0 &  \(0.600624-0.092598\imag\) &  \(0.599471-0.092695\imag\) &  \(0.599444-0.092702\imag\) &  \(0.599443-0.092703\imag\) \\
 3 &  1 &  \(0.583963-0.280957\imag\) &  \(0.582676-0.281267\imag\) &  \(0.582643-0.281289\imag\) &  \(0.582642-0.281290\imag\) \\
 3 &  2 &  \(0.553183-0.478417\imag\) &  \(0.551639-0.479004\imag\) &  \(0.551595-0.479045\imag\) &  \(0.551594-0.479047\imag\) \\
 3 &  3 &  \(0.513059-0.689431\imag\) &  \(0.511160-0.690400\imag\) &  \(0.511101-0.690465\imag\) &  \(0.511099-0.690468\imag\) \\
 4 &  0 &  \(0.810671-0.094060\imag\) &  \(0.809213-0.094157\imag\) &  \(0.809179-0.094164\imag\) &  \(0.809178-0.094164\imag\) \\
 4 &  1 &  \(0.798221-0.284008\imag\) &  \(0.796670-0.284311\imag\) &  \(0.796632-0.284333\imag\) &  \(0.796631-0.284334\imag\) \\
 4 &  2 &  \(0.774473-0.479311\imag\) &  \(0.772741-0.479859\imag\) &  \(0.772697-0.479898\imag\) &  \(0.772695-0.479900\imag\) \\
 4 &  3 &  \(0.741708-0.682982\imag\) &  \(0.739721-0.683841\imag\) &  \(0.739667-0.683899\imag\) &  \(0.739665-0.683902\imag\) \\
 4 &  4 &  \(0.703005-0.897121\imag\) &  \(0.700708-0.898375\imag\) &  \(0.700642-0.898456\imag\) &  \(0.700641-0.898460\imag\) \\
\bottomrule
\end{tabular}
\caption{\label{tab:qnms-gravitational-field}Quasinormal frequencies of gravitational perturbations for \(p = 3\) and \(q > 1\). for the regular Black Hole where \(r_0/M_0 = 0.1\) and .}
\end{table}

\begin{figure}[htb]
\centering
\includegraphics[width=0.95\linewidth]{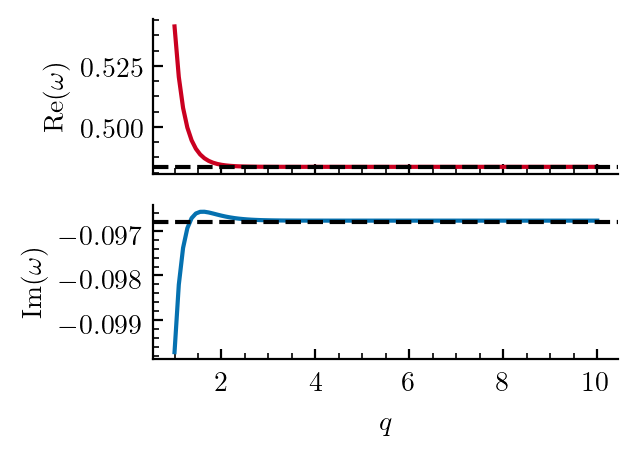}
\caption{\label{fig:scalar-qnm}Real and imaginary parts of the quasinormal mode \((\ell, n)=(2,0)\) of scalar perturbations for \(p = 3\) and \(q > 0\). The dashed lines correspond to the respective values of the real and imaginary parts of the Classical Schwarzschild black hole.}
\end{figure}

\begin{figure}[htb]
\centering
\includegraphics[width=0.95\linewidth]{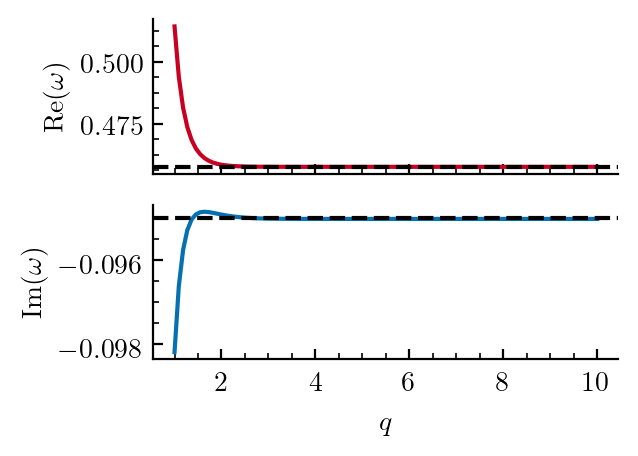}
\caption{\label{fig:vector-qnm}Real and imaginary parts of the quasinormal mode \((\ell, n)=(2,0)\) of electromagnetic perturbations for \(p = 3\) and \(q > 0\). The dashed lines correspond to the respective values of the real and imaginary parts of the Classical Schwarzschild black hole.}
\end{figure}

\begin{figure}[htb]
\centering
\includegraphics[width=0.95\linewidth]{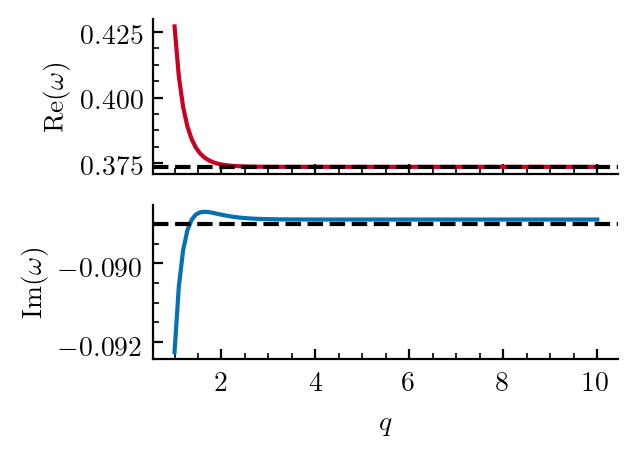}
\caption{\label{fig:tensor-qnm}Real and imaginary parts of the quasinormal mode \((\ell, n)=(2,0)\) of gravitational perturbations for \(p = 3\) and \(q > 0\). The dashed lines correspond to the respective values of the real and imaginary parts of the Classical Schwarzschild black hole.}
\end{figure}

We can notice that all QNMs possess a negative imaginary part, conferring therefore stability for the black hole. Also, both real and imaginary parts of the QNMs do not seem to be wildly sensible to the parameter \(q\) in the limit \(q\to\infty\), indicating that the damping time is, with good approximation, independent of it. However, we can observe a change the case \(q\leq 1\), when the BH is smaller damping time than the classical case. 

\section{Absorption cross section}
\label{sec:gbf-computation}

Another interesting aspect for investigate the field perturbations around a black hole spacetime is the amount of plunging field which is absorbed by the black hole, the absorption cross section. It embodies the likelihood of a particle to be scattered/deflected by the black hole. Based on the quantum mechanics analogy, the cross-section can be written as
\begin{equation}
\label{eq:total-cross-section}
\sigma_s(\omega) = \sum_{\ell = 0}^{\infty} \sigma^{(\ell)}_{s}(\omega),
\end{equation}
where we defined the partial absorption cross section
\begin{equation}
\label{eq:partial-cross-section}
\sigma^{(\ell)}_{s}(\omega) = \frac{\pi}{\omega^{2}} (2\ell + 1) T_{\ell,s}(\omega),
\end{equation}
where \(T_{\ell,s}(\omega)\) is the transmission coefficient. To compute the transmission coefficient we employ the transfer matrix method, as described in~\cite{Du_1996}.

\begin{figure}[htb]
\centering
\includegraphics[width=0.95\linewidth]{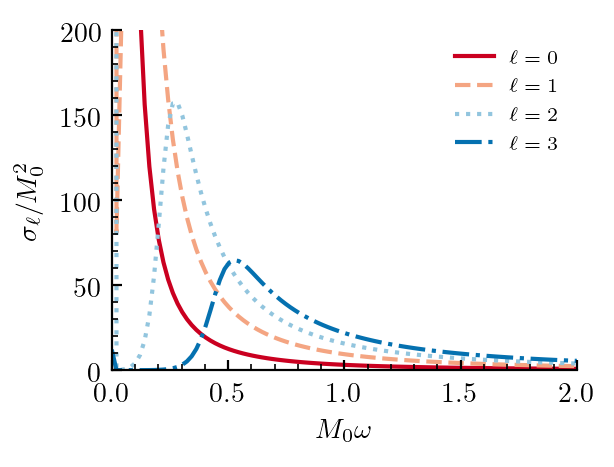}
\caption{\label{fig:partial-absorption-cross-section}The partial absorption cross section for a massless scalar wave impinging upon a Bardeen Black Hole (\(p = 3, q = 2\)) for the first six multipoles for \(r_0/M_0 = 0.1\).}
\end{figure}

The partial absorption cross sections versus \(M_0 \omega\) for \(\ell = 0\) to \(\ell=3\) are plotted in Fig. \ref{fig:partial-absorption-cross-section}. One can notice that the \(\ell=0\) contribution is the largest, responsible for the non-vanishing of the cross section in the small energy limit. We observe how \(\sigma_{\ell}/M_0^2\) decreases when the value of \(\ell\) increases, due to the fact that the effective potential peak increases for larger values of \(\ell\).

\section{Conclusion}
\label{sec:conclusion}

In the present paper we have investigated the quasinormal frequencies and the greybody factors of gravitational perturbations around a class of regular black hole solutions. To obtain the corresponding quasinormal frequencies, we perturbed the background and taking advantage of the symmetry we are able to decompose the original function into spherical harmonics, writing down the differential equation of the problem involved. Then we consider the sixth order WKB approximation to obtain the QN frequencies. Further, as can be observed in tables, our solutions reveal that the family of black holes here analysed recover the classical behaviour for \(q \to \infty\), and possess a stronger absorption section in the microscopic regime. That means that BH regularity amends the dissipative effect of the black hole on its neighborhood. Finally, as can be observed in tables, for \(q > 1\), the imaginary part of the quasinormal frequencies are also negative. Thus, our results indicate that all modes are found to be unstable.

\section{Acknowledgments}
\label{sec:acknowledgments}
The author V. S. would like to thank the Fundação Cearense de apoio ao Desenvolvimento Científico e Tecnológico (FUNCAP) for financial support.
The author A. R. acknowledges DI-VRIEA for financial support through Proyecto Postdoctorado 2019 VRIEA-PUCV.

\bibliographystyle{spphys}
\bibliography{gbf}

\begin{thebibliography}{10}
\providecommand{\url}[1]{{#1}}
\providecommand{\urlprefix}{URL }
\expandafter\ifx\csname urlstyle\endcsname\relax
  \providecommand{\doi}[1]{DOI \discretionary{}{}{}#1}\else
  \providecommand{\doi}{DOI \discretionary{}{}{}\begingroup
  \urlstyle{rm}\Url}\fi

\bibitem{Einstein:1916vd}
A.~Einstein, Annalen Phys. \textbf{49}(7), 769 (1916).
\newblock \doi{10.1002/andp.200590044, 10.1002/andp.19163540702}.
\newblock [Annalen Phys.14,517(2005); ,65(1916); Annalen
  Phys.354,no.7,769(1916)]

\bibitem{israel_event_1967}
W.~Israel, Physical Review \textbf{164}(5), 1776 (1967).
\newblock \doi{10.1103/PhysRev.164.1776}.
\newblock \urlprefix\url{https://link.aps.org/doi/10.1103/PhysRev.164.1776}.
\newblock Publisher: American Physical Society

\bibitem{israel_event_1968}
W.~Israel, Communications in Mathematical Physics \textbf{8}(3), 245 (1968).
\newblock \doi{10.1007/BF01645859}.
\newblock \urlprefix\url{https://doi.org/10.1007/BF01645859}

\bibitem{carter_axisymmetric_1971}
B.~Carter, Physical Review Letters \textbf{26}(6), 331 (1971).
\newblock \doi{10.1103/PhysRevLett.26.331}.
\newblock \urlprefix\url{https://link.aps.org/doi/10.1103/PhysRevLett.26.331}.
\newblock Publisher: American Physical Society

\bibitem{Hawking:1974rv}
S.W. Hawking, Nature \textbf{248}, 30 (1974).
\newblock \doi{10.1038/248030a0}

\bibitem{Hawking:1974sw}
S.W. Hawking, Commun. Math. Phys. \textbf{43}, 199 (1975).
\newblock \doi{10.1007/BF02345020, 10.1007/BF01608497}.
\newblock [,167(1975)]

\bibitem{Kanti:2002nr}
P.~Kanti, J.~March-Russell, Phys. Rev. D \textbf{66}, 024023 (2002).
\newblock \doi{10.1103/PhysRevD.66.024023}

\bibitem{PhysRevD.93.024045}
B.~Carneiro~da Cunha, F.~Novaes, Phys. Rev. D \textbf{93}, 024045 (2016).
\newblock \doi{10.1103/PhysRevD.93.024045}.
\newblock \urlprefix\url{https://link.aps.org/doi/10.1103/PhysRevD.93.024045}

\bibitem{Castro_2013}
A.~Castro, J.M. Lapan, A.~Maloney, M.J. Rodriguez, Classical and Quantum
  Gravity \textbf{30}(16), 165005 (2013).
\newblock \doi{10.1088/0264-9381/30/16/165005}.
\newblock
  \urlprefix\url{https://doi.org/10.1088%2F0264-9381%2F30%2F16%2F165005}

\bibitem{Gursel:2019fyd}
H.~Gürsel, I.~Sakall\i, Eur. Phys. J. C \textbf{80}(3), 234 (2020).
\newblock \doi{10.1140/epjc/s10052-020-7791-3}

\bibitem{Hyun:2019qgq}
Y.H. Hyun, Y.~Kim, S.C. Park, JHEP \textbf{06}, 041 (2019).
\newblock \doi{10.1007/JHEP06(2019)041}

\bibitem{Kanzi:2019gtu}
S.~Kanzi, I.~Sakall\i, Nucl. Phys. B \textbf{946}, 114703 (2019).
\newblock \doi{10.1016/j.nuclphysb.2019.114703}

\bibitem{Chowdhury:2020bdi}
A.~Chowdhury, N.~Banerjee, Phys. Lett. B \textbf{805}, 135417 (2020).
\newblock \doi{10.1016/j.physletb.2020.135417}

\bibitem{Gray:2015xig}
F.~Gray, M.~Visser, Universe \textbf{4}(9), 93 (2018).
\newblock \doi{10.3390/universe4090093}

\bibitem{AyonBeato:1998ub}
E.~Ayon-Beato, A.~Garcia, Phys. Rev. Lett. \textbf{80}, 5056 (1998).
\newblock \doi{10.1103/PhysRevLett.80.5056}

\bibitem{Kokkotas_Schmidt_1999}
K.D. Kokkotas, B.G. Schmidt, Living Reviews in Relativity \textbf{2}(1), 2
  (1999).
\newblock \doi{10.12942/lrr-1999-2}

\bibitem{Abbott:2016blz}
B.P. Abbott, et~al., Phys. Rev. Lett. \textbf{116}(6), 061102 (2016).
\newblock \doi{10.1103/PhysRevLett.116.061102}

\bibitem{Akiyama:2019cqa}
K.~Akiyama, et~al., Astrophys. J. \textbf{875}(1), L1 (2019).
\newblock \doi{10.3847/2041-8213/ab0ec7}

\bibitem{Akiyama:2019eap}
K.~Akiyama, et~al., Astrophys. J. \textbf{875}(1), L6 (2019).
\newblock \doi{10.3847/2041-8213/ab1141}

\bibitem{Konoplya_Zhidenko_2011}
R.A. Konoplya, A.~Zhidenko, Reviews of Modern Physics \textbf{83}(3), 793–836
  (2011).
\newblock \doi{10.1103/RevModPhys.83.793}

\bibitem{Kokkotas_Schutz_1992}
K.D. Kokkotas, B.F. Schutz, Monthly Notices of the Royal Astronomical Society
  \textbf{255}(1), 119–128 (1992).
\newblock \doi{10.1093/mnras/255.1.119}

\bibitem{1971ApJ..170L.105P}
W.H. {Press}, apjl \textbf{170}, L105 (1971).
\newblock \doi{10.1086/180849}

\bibitem{Vishveshwara_1970}
C.V. Vishveshwara, Nature \textbf{227}(52615261), 936–938 (1970).
\newblock \doi{10.1038/227936a0}

\bibitem{Giesler_Isi_Scheel_Teukolsky_2019}
M.~Giesler, M.~Isi, M.A. Scheel, S.A. Teukolsky, Physical Review X
  \textbf{9}(4), 041060 (2019).
\newblock \doi{10.1103/PhysRevX.9.041060}

\bibitem{Flachi:2012nv}
A.~Flachi, J.P. Lemos, Phys. Rev. D \textbf{87}(2), 024034 (2013).
\newblock \doi{10.1103/PhysRevD.87.024034}

\bibitem{Kokkotas:1999bd}
K.D. Kokkotas, B.G. Schmidt, Living Rev. Rel. \textbf{2}, 2 (1999).
\newblock \doi{10.12942/lrr-1999-2}

\bibitem{Konoplya:2003ii}
R.~Konoplya, Phys. Rev. D \textbf{68}, 024018 (2003).
\newblock \doi{10.1103/PhysRevD.68.024018}

\bibitem{Konoplya:2011qq}
R.~Konoplya, A.~Zhidenko, Rev. Mod. Phys. \textbf{83}, 793 (2011).
\newblock \doi{10.1103/RevModPhys.83.793}

\bibitem{Barrau:2019swg}
A.~Barrau, K.~Martineau, J.~Martinon, F.~Moulin, Phys. Lett. B \textbf{795},
  346 (2019).
\newblock \doi{10.1016/j.physletb.2019.06.033}

\bibitem{Konoplya:2019hlu}
R.~Konoplya, A.~Zhidenko, A.~Zinhailo, Class. Quant. Grav. \textbf{36}, 155002
  (2019).
\newblock \doi{10.1088/1361-6382/ab2e25}

\bibitem{Panotopoulos:2017hns}
G.~Panotopoulos, A.~Rinc\'{o}ón, Int. J. Mod. Phys. D \textbf{27}(03), 1850034
  (2017).
\newblock \doi{10.1142/S0218271818500347}

\bibitem{Rincon:2018sgd}
A.~Rinc\'{o}n, G.~Panotopoulos, Phys. Rev. D \textbf{97}(2), 024027 (2018).
\newblock \doi{10.1103/PhysRevD.97.024027}

\bibitem{Destounis:2018utr}
K.~Destounis, G.~Panotopoulos, A.~Rinc\'{o}n, Eur. Phys. J. C \textbf{78}(2),
  139 (2018).
\newblock \doi{10.1140/epjc/s10052-018-5576-8}

\bibitem{Rincon:2018ktz}
A.~Rinc\'{o}n, G.~Panotopoulos, Eur. Phys. J. C \textbf{78}(10), 858 (2018).
\newblock \doi{10.1140/epjc/s10052-018-6352-5}

\bibitem{Panotopoulos:2019qjk}
G.~Panotopoulos, A.~Rinc\'{o}n, Eur. Phys. J. Plus \textbf{134}(6), 300 (2019).
\newblock \doi{10.1140/epjp/i2019-12686-x}

\bibitem{Rincon:2020iwy}
A.~Rinc\'{o}n, G.~Panotopoulos, Phys. Dark Univ. \textbf{30}, 100639 (2020).
\newblock \doi{10.1016/j.dark.2020.100639}

\bibitem{Rincon:2020pne}
A.~Rincon, G.~Panotopoulos.
\newblock {Quasinormal modes of black holes with a scalar hair in
  Einstein-Maxwell-dilaton theory} (2020)

\bibitem{Panotopoulos:2020zbj}
G.~Panotopoulos.
\newblock {Quasinormal modes of charged black holes in higher-dimensional
  Einstein-power-Maxwell theory} (2020).
\newblock \doi{10.3390/axioms9010033}

\bibitem{Oliveira:2018oha}
R.~Oliveira, D.~Dantas, V.~Santos, C.~Almeida, Class. Quant. Grav.
  \textbf{36}(10), 105013 (2019).
\newblock \doi{10.1088/1361-6382/ab1873}

\bibitem{Santos:2015gja}
V.~Santos, R.~Maluf, C.~Almeida, Phys. Rev. D \textbf{93}(8), 084047 (2016).
\newblock \doi{10.1103/PhysRevD.93.084047}

\bibitem{Devi:2020uac}
S.~Devi, R.~Roy, S.~Chakrabarti, Eur. Phys. J. C \textbf{80}(8), 760 (2020).
\newblock \doi{10.1140/epjc/s10052-020-8311-1}

\bibitem{Jusufi:2020agr}
K.~Jusufi, M.~Amir, M.S. Ali, S.D. Maharaj.
\newblock {Quasinormal modes, shadow and greybody factors of 5D electrically
  charged Bardeen black holes} (2020)

\bibitem{doi:10.1142/S021827181641008X}
C.F.B. Macedo, L.C.B. Crispino, E.S. de~Oliveira, International Journal of
  Modern Physics D \textbf{25}(09), 1641008 (2016).
\newblock \doi{10.1142/S021827181641008X}.
\newblock \urlprefix\url{https://doi.org/10.1142/S021827181641008X}

\bibitem{hawking1975large}
S.W. Hawking, G.F.R. Ellis, \emph{The Large Scale Structure of Space-Time
  (Cambridge Monographs on Mathematical Physics)} (Cambridge University Press,
  1975).
\newblock
  \urlprefix\url{http://www.amazon.com/Structure-Space-Time-Cambridge-Monographs-Mathematical/dp/0521099064}

\bibitem{Penrose_2002}
R.~Penrose, General Relativity and Gravitation \textbf{34}(7), 1141–1165
  (2002).
\newblock \doi{10.1023/A:1016578408204}

\bibitem{Frolov:1979tu}
V.P. Frolov, G.~Vilkovisky, in \emph{{The Second Marcel Grossmann Meeting on
  the Recent Developments of General Relativity (In Honor of Albert Einstein)}}
  (1979), p. 0455

\bibitem{Hayward_2006}
S.A. Hayward, Physical Review Letters \textbf{96}(3), 031103 (2006).
\newblock \doi{10.1103/PhysRevLett.96.031103}

\bibitem{Toshmatov:2015wga}
B.~Toshmatov, A.~Abdujabbarov, Z.e. Stuchlík, B.~Ahmedov, Phys. Rev. D
  \textbf{91}(8), 083008 (2015).
\newblock \doi{10.1103/PhysRevD.91.083008}

\bibitem{Fernando_Correa_2012}
S.~Fernando, J.~Correa, Physical Review D \textbf{86}(6), 064039 (2012).
\newblock \doi{10.1103/PhysRevD.86.064039}

\bibitem{Flachi_Lemos_2013}
A.~Flachi, J.P.S. Lemos, Physical Review D \textbf{87}(2), 024034 (2013).
\newblock \doi{10.1103/PhysRevD.87.024034}

\bibitem{Toshmatov_Abdujabbarov_Stuchlik_Ahmedov_2015}
B.~Toshmatov, A.~Abdujabbarov, Z.~Stuchl\'{i}k, B.~Ahmedov, Physical Review D
  \textbf{91}(8), 083008 (2015).
\newblock \doi{10.1103/PhysRevD.91.083008}

\bibitem{Toshmatov_Stuchlik_Schee_Ahmedov_2018}
B.~Toshmatov, Z.~Stuchl\'{i}k, J.~Schee, B.~Ahmedov, Physical Review D
  \textbf{97}(8), 084058 (2018).
\newblock \doi{10.1103/PhysRevD.97.084058}

\bibitem{Panotopoulos:2018pvu}
G.~Panotopoulos, A.~Rincón, Phys. Rev. D \textbf{97}(8), 085014 (2018).
\newblock \doi{10.1103/PhysRevD.97.085014}

\bibitem{Dey:2018cws}
S.~Dey, S.~Chakrabarti, Eur. Phys. J. C \textbf{79}(6), 504 (2019).
\newblock \doi{10.1140/epjc/s10052-019-7004-0}

\bibitem{Wu:2017tfo}
C.~Wu.
\newblock {Quasinormal modes of gravitational perturbation around some
  well-known regular black holes} (2017)

\bibitem{Yekta:2019por}
D.~Mahdavian~Yekta, M.~Karimabadi, S.~Alavi.
\newblock {Quasinormal modes of Regular black holes: Non-minimally coupled
  massive scalar field perturbations} (2019)

\bibitem{Neves_Saa_2014}
J.C.S. Neves, A.~Saa, Physics Letters B \textbf{734}, 44–48 (2014).
\newblock \doi{10.1016/j.physletb.2014.05.026}

\bibitem{ansoldi2008spherical}
S.~Ansoldi.
\newblock Spherical black holes with regular center: a review of existing
  models including a recent realization with gaussian sources (2008)

\bibitem{PhysRevD.35.3632}
S.~Iyer, Phys. Rev. D \textbf{35}, 3632 (1987).
\newblock \doi{10.1103/PhysRevD.35.3632}.
\newblock \urlprefix\url{https://link.aps.org/doi/10.1103/PhysRevD.35.3632}

\bibitem{1985ApJ291L33S}
B.F. {Schutz}, C.M. {Will}, Astrophysical Journal, Letters \textbf{291}, L33
  (1985).
\newblock \doi{10.1086/184453}

\bibitem{PhysRevD.35.3621}
S.~Iyer, C.M. Will, Phys. Rev. D \textbf{35}, 3621 (1987).
\newblock \doi{10.1103/PhysRevD.35.3621}.
\newblock \urlprefix\url{https://link.aps.org/doi/10.1103/PhysRevD.35.3621}

\bibitem{PhysRevD.68.024018}
R.A. Konoplya, Phys. Rev. D \textbf{68}, 024018 (2003).
\newblock \doi{10.1103/PhysRevD.68.024018}.
\newblock \urlprefix\url{https://link.aps.org/doi/10.1103/PhysRevD.68.024018}

\bibitem{Du_1996}
M.L. Du, Communications in Theoretical Physics \textbf{25}(3), 257–260
  (1996).
\newblock \doi{10.1088/0253-6102/25/3/257}

\end{thebibliography}

\end{document}